\pdfoutput=1
\documentclass[letterpaper,twoside,journal]{IEEEtran}

\makeatletter
\let\NAT@parse\undefined
\makeatother
\usepackage[numbers,sort&compress]{natbib}

\usepackage{mathtools}
\usepackage{amssymb}
\usepackage{algorithm,algpseudocode}
\usepackage{multicol}
\usepackage{balance} %

\usepackage[table]{xcolor}
\usepackage{pgf}
\usepackage{mathrsfs}
\usepackage{tabularx}
\usepackage{amsmath}
\usepackage{multirow, makecell}
\usepackage{wrapfig}
\usepackage{dsfont}
\usepackage{soul}

\usepackage{bm}      %
\usepackage{mathtools}
\usepackage{xspace}
\usepackage{booktabs}
\usepackage{hyperref}
\usepackage{subcaption}
\usepackage{etoolbox} %
\usepackage{eso-pic} 

\newcommand{\rotatedHeader}[3][]{\multirow[t]{#2}{*}[#1]{\rotatebox[origin=c]{90}{\footnotesize #3}}}
\newcolumntype{N}{r@{\hspace{.7mm}}c@{\hspace{.7mm}}l}

\newcommand{\E}[2]{\Diamond_{[#1,#2]}}
\newcommand{\G}[2]{\square_{[#1,#2]}}
\newcommand{\U}[2]{\ \mathcal{U}_{[#1,#2]}}

\definecolor{airforceblue}{rgb}{0.36, 0.54, 0.66}
\definecolor{applegreen}{rgb}{0.55, 0.71, 0.0}
\definecolor{ballblue}{rgb}{0.13, 0.67, 0.8}

\definecolor{tab_blue}{rgb}{0.0, 0.5, 1.0}
\definecolor{tab_green}{rgb}{0.4, 0.69, 0.2}
\definecolor{pgd}{rgb}{0.0, 0.5, 1.0}
\definecolor{pgd_comp}{RGB}{175, 255, 255}
\sethlcolor{pgd_comp}

\newcommand{\traj}{\tau}
\newcommand{\goaltraj}{\tau_g}  %
\newcommand{\goaltraji}[1]{\tau_{g_{#1}}}  %
\newcommand{\statespace}{S}  

\newcommand{\R}{\mathds{R}}

\newcolumntype{?}{!{\vrule width 1pt}}

\newcommand\seq{\textit{seq}}

\newcommand\cover{\textit{cover}}
\newcommand\branch{\textit{branch}}
\newcommand\loopspec{\textit{loop}}
\newcommand\signalspec{\textit{signal}}
\newcommand\mixedspec{\textit{Mixed}}
\newcommand\Seq{\textit{Seq.}}

\newcommand\Cover{\textit{Cover}}
\newcommand\Branch{\textit{Branch}}
\newcommand\Loopspec{\textit{Loop}}

\newcommand\Mixedspec{\textit{Mixed}}
\newcommand\teamchoice{\textit{Choice-Seq}}
\newcommand\teamredundant{\textit{Redundant}}

\newtheorem{definition}{Definition}

\newcommand{\quotes}[1]{``#1''}

\newcommand{\nnnum}[1]{\relax\ifmmode 
	{\mathbb #1}_{\geq 0} \else ${\mathbb #1}_{\geq 0}$
	\fi}
\newcommand{\npnum}[1]{\relax\ifmmode 
	{\mathbb #1}_{\leq 0} \else ${\mathbb #1}_{\leq 0}$
	\fi}
\newcommand{\pnum}[1]{\relax\ifmmode 
	{\mathbb #1}_{> 0} \else ${\mathbb #1}_{> 0}$
	\fi}
\newcommand{\nnum}[1]{\relax\ifmmode 
	{\mathbb #1}_{< 0} \else ${\mathbb #1}_{< 0}$
	\fi}
\newcommand{\plnum}[1]{\relax\ifmmode 
	{\mathbb #1}_{+} \else ${\mathbb #1}_{+}$
	\fi}
\newcommand{\nenum}[1]{\relax\ifmmode 
	{\mathbb #1}_{-} \else ${\mathbb #1}_{-}$
	\fi}

\newcommand{\Time}{{\num T}}

\newcommand\reallywidehat[1]{%
	\savestack{\tmpbox}{\stretchto{%
			\scaleto{%
				\scalerel*[\widthof{\ensuremath{#1}}]{\kern-.6pt\bigwedge\kern-.6pt}%
				{\rule[-\textheight/2]{1ex}{\textheight}}%
			}{\textheight}%
		}{0.5ex}}%
	\stackon[1pt]{#1}{\tmpbox}%
}
\newcommand{\AlignedComment}[1]{ \text{// #1}}

\newcommand{\norm}[1]{\left\lVert#1\right\rVert}

\newcommand{\mname}[1]{\textsc{#1}} 
\newcommand{\gcbfp}{GCBF+}
\newcommand{\gcbf}{GCBF}
\newcommand{\oldmastl}{MA-STL}
\newcommand{\mastl}{\oldmastl-A}
\newcommand{\teamspecs}{CaTL+}
\newcommand{\catltask}[2]{\langle #1, #2 \rangle}
\newcommand{\ctask}{\mathcal{T}}
\newcommand{\taskcount}{m}
\newcommand{\stlpy}{\mname{STLPY-SA}}
\newcommand{\stlpyglobal}{\mname{STLPY-Global}}

\newcommand{\pwlmastl}{\mname{PWL MA-STL}}
\newcommand{\cenlplan}{\mname{Gradient}}
\newcommand{\maspec}{\Psi}

\newcommand{\fullmaspec}{\bigwedge_{i=1}^N \phi_i}
\newcommand{\agents}{\mathcal{N}}
\newcommand{\safeset}{\mathcal{S}_s}
\newcommand{\unsafeset}{\mathcal{S}_u}
\newcommand{\plannn}[2][]{\pi^{\phi\ifstrempty{#2}{}{_{#2}}}_g\ifstrempty{#1}{}{(#1)}} %
\newcommand{\jointplannn}[1][]{\pi^{\psi}_g\ifstrempty{#1}{}{(#1)}} %
\newcommand{\sbar}[1][]{\ifstrempty{#1}{\bar{s}}{\bar{s}(#1)}}
\newcommand{\goal}[2][]{g_{#2}\ifstrempty{#1}{}{(#1)}}

\newcommand{\statevar}[2][]{s_{#2}\ifstrempty{#1}{}{(#1)}}
\newcommand{\action}[2][]{u_{#2}\ifstrempty{#1}{}{(#1)}}
\newcommand{\positionvar}[2][]{p_{#2}\ifstrempty{#1}{}{(#1)}}
\newcommand{\positionspace}{\mathbb{P}}
\newcommand{\filterstate}[1][]{\texttt{filter}_{\positionvar{i}}\ifstrempty{#1}{}{(#1)}}
\newcommand{\lossstl}{\mathcal{L}_{\text{STL}}}
\newcommand{\lossacheivable}{\mathcal{L}_{\text{ach}}}

\newcommand{\coeffstl}{\lambda_{\text{STL}}}
\newcommand{\coeffacheivable}{\lambda_{\text{ach}}}

\newcommand{\trajdist}[2]{D_{#2}\ifstrempty{#1}{}{(#1)}}
\newcommand{\graph}[1][]{G\ifstrempty{#1}{}{(#1)}}
\newcommand{\vertices}[1][]{V\ifstrempty{#1}{}{(#1)}}
\newcommand{\edges}[1][]{E\ifstrempty{#1}{}{(#1)}}
\newcommand{\vertice}[2][]{v_{#2}\ifstrempty{#1}{}{(#1)}}
\newcommand{\planner}{\mname{GNN-ODE}}

\newcommand{\divAPC}{Agents per Cluster}%
\newcommand{\divPO}{Path Overlap}%

\newcommand{\ndiff}{N_{\text{diff}}}
\newcommand{\diffplanner}{\mname{Diff-MA}}

\newcommand{\diffsaplanner}{\mname{Diff-SA}}
\newcommand{\guidancefn}{f_{\text{guid}}}
\newcommand{\dtau}{d_{\tau}}  %
\newcommand{\nachfrac}{k_{\text{ach}}}  %
\newcommand{\epsresample}{\epsilon_{\text{sample}}}  %
\newcommand{\sigmax}{\sigma_{\text{max}}}  %
\newcommand{\sigmin}{\sigma_{\text{min}}}  %

\newcommand{\taustate}[2][]{\ifstrempty{#1}{\boldsymbol{{\tau}_{s_{#2}}}}{\boldsymbol{{\tau}^{#1}_{s_{#2}}}}}
\newcommand{\taudiff}[2][]{\ifstrempty{#1}{\boldsymbol{{\tau}_{g_{#2}}}}{\boldsymbol{{\tau}^{#1}_{g_{#2}}}}}
\newcommand{\tautilde}[2][]{\ifstrempty{#1}{\boldsymbol{\tilde{\tau}_{g_{#2}}}}{\boldsymbol{\tilde{\tau}^{#1}_{g_{#2}}}}}
\newcommand{\bartautilde}[1][]{\ifstrempty{#1}{\boldsymbol{\tilde{{\tau}}_{g}}}{\boldsymbol{\tilde{{\tau}}^{#1}_{g}}}}
\newcommand{\tauhat}[2][]{\ifstrempty{#1}{\boldsymbol{\hat{\tau}_{g_{#2}}}}{\boldsymbol{\hat{\tau}^{#1}_{g_{#2}}}}}

\newcommand{\btau}[1][]{\ifstrempty{#1}{\boldsymbol{{\tau}}}{\boldsymbol{{\tau}}^{#1}}}
\newcommand{\btautilde}[1][]{\ifstrempty{#1}{\boldsymbol{\tilde{\tau}}}{\boldsymbol{\tilde{\tau}}^{#1}}}
\newcommand{\btauhat}[1][]{\ifstrempty{#1}{\boldsymbol{\hat{\tau}}}{\boldsymbol{\hat{\tau}}^{#1}}}

\newcommand{\maxresample}{N_{\text{sample}}}

\newcommand{\Predset}{\mathscr{P}}

\newcommand{\websiteurl}{https://www.jeappen.com/diff-ma-stl/}

\newcommand{\tWedge}{\mathop{\textstyle\bigwedge}}
\newcommand{\tVee}{\mathop{\textstyle\bigvee}}

\graphicspath{ {./images/} } 

\title{Generalizable Multi-Agent Planning from Signal Temporal Logic Specifications via Diffusion} %

\author{
    Joe Eappen, Zikang Xiong,  Shreyash S. Iyengar, and Suresh Jagannathan %
    \thanks{Manuscript received: April 21, 2026; Revised: July 15, 2026; Accepted: August 7, 2026.}%
    \thanks{This paper was recommended for publication by Editor M. Ani Hsieh upon evaluation of the Associate Editor and Reviewers' comments.} %
    \thanks{ The authors are affiliated with the ECE and CS Departments of Purdue University. Contact: \texttt{\{jeappen,sjaganna\}@purdue.edu}} %
    \thanks{Digital Object Identifier (DOI): see top of this page.}
}

\begin{document}
\bstctlcite{RAL_Ref_Control}
\AddToShipoutPictureBG*{\AtPageLowerLeft{%
  \hspace*{\dimexpr 1in+\oddsidemargin\relax}%
  \raisebox{12pt}{\parbox[b]{\textwidth}{\scriptsize
    \textcopyright~2026 IEEE. Personal use of this material is permitted. Permission from IEEE must be obtained for all other uses, in any current or future media, including reprinting/republishing this material for advertising or promotional purposes, creating new collective works, for resale or redistribution to servers or lists, or reuse of any copyrighted component of this work in other works.}}}}
\maketitle

\begin{abstract}    
    Multi‑agent systems in the real-world (e.g., drone swarms, autonomous cars, warehouse robots) must satisfy rich, temporal tasks while avoiding collisions. Signal Temporal Logic (STL) elegantly encodes such objectives, but current STL planning methods face critical limitations.
    State-of-the-art optimization-based approaches can handle arbitrary STL specifications but struggle with scalability, becoming computationally impractical as the number of agents grows. 
    Learning-based methods efficiently handle a large number of agents with rapid planning times but fare poorly when deployment-time objectives differ from those used during training, and do not support planning tasks that require different specifications to be ascribed to different agents (i.e., heterogeneity) or team-level specifications requiring coordination of multiple agents. 
    This fundamental trade-off between \emph{generalizability} and \emph{scalability} presents a challenge for realizing multi-agent STL planning algorithms in practice.
    To overcome this challenge, we introduce a new diffusion method for multi-agent planning with STL specifications. 
    Using a differentiable approximation of STL, we integrate the STL gradient in the denoising process, making our approach generalizable to novel formulas whose predicates are placed anywhere within the goal region covered during training, while achieving the same scalability as existing learning-based methods. 
    Our method supports  
    heterogeneous specifications, and by using diffusion models, naturally enhances plan \emph{diversity}, thereby significantly reducing safety-related violations (e.g., collisions) among agents. A detailed evaluation study justifies the utility of STL-guided diffusion-based multi-agent planners for constructing generalizable, scalable, and diverse plans.
    Videos and code are available at \url{https://www.jeappen.com/diff-ma-stl/} and \url{https://github.com/jeappen/diff-ma-stl}.
\end{abstract}


\newcommand{\BibTeX}{\rm B\kern-.05em{\sc i\kern-.025em b}\kern-.08em\TeX}

\section{Introduction}

\label{sec:intro}
Symbolic specification methods have recently experienced a resurgence through neuro-symbolic algorithms \citep{Chaudhuri2021},
 which aim to combine the generalizability of neural methods with the interpretability and modifiability of symbolic systems by human users.
 Symbolic techniques have also emerged in robot motion planning, particularly with the use of Signal Temporal Logic (STL) to specify objectives for multi-robot systems, which can then be addressed by Mixed-Integer Linear Programming (MILP) solvers \citep{Sun2022}, graph-based algorithms \citep{Buyukkocak2021}, or sampling-based methods \citep{Vasile2020}.
 STL specifies complex temporal properties over continuous signals, making it suitable for complex temporal tasks in robotics \citep{pmlr-v155-puranic21a} such as coverage, patrolling, and surveillance \citep{wang2024tractable}.
Importantly, collision avoidance remains a practical challenge for multi-robot systems operating in crowded spaces under complex logic constraints \citep{qin2021learning, zhang_gcbf_2024}. At the same time, prominent existing methods \citep{Sun2022} have been shown to scale poorly as the number of agents and the complexity of the specifications increase \citep{eappen2024scaling}.

\begin{figure}[t!]
	\centering
	\includegraphics[width=\linewidth]{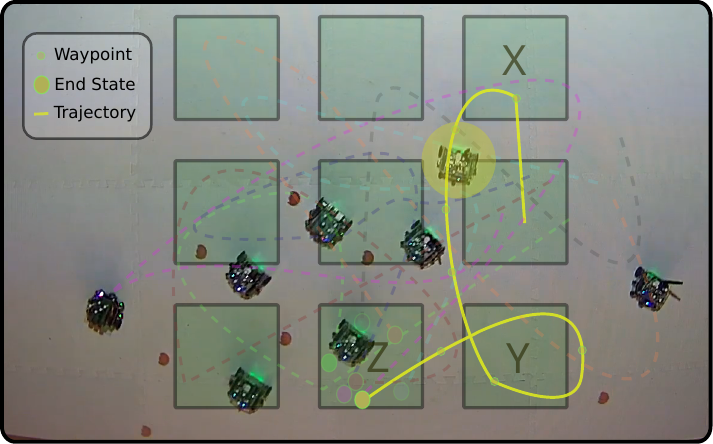}
	\caption{\small
		Demo on $N=10$ differential drive robots \citep{CSS:robotarium}
		following a (heterogeneous) \seq~ specification (Sec. \ref{sec:exp-setup-specs}) that requires agents to visit designated locations in a prescribed order (X, Y, Z in the figure for the \textcolor{yellow!60!black}{yellow} robot) at specified time intervals. A video is provided on our website: \href{\websiteurl}{\websiteurl}
	}
	\label{fig:robot_demo}
\end{figure}
\begin{table}[t!]
\centering
\captionsetup{labelfont={small, bf}, textfont=small}
\small %
\setlength{\tabcolsep}{2pt} %

\caption{\small A comparison of \mastl~planner choices. 
Our diffusion-based method excels in Diversity and Generalizability with capabilities beyond other planners. \textbf{Het. (Specs)}: Heterogeneous Specifications i.e., different tasks among agents, \textbf{MA Aware}: Considers Multi-agent interactions while planning, 
  \textbf{Cmplx. Specs}: Can handle complex specifications (with 
  $N>5$ 
  agents as studied in prior work \citep[Sec.~3]{eappen2024scaling}), \textbf{Scale}: Safely scale to large $N$ (at least 32 agents),
\textbf{Plan. Time}: Average planning time (s) to generate a single end-to-end plan for novel agent placements, \textbf{Div.}: Exhibits Plan Diversity (as in Sec~\ref{subsec:diversity}), 
\textbf{Gen.}: Allows Generalizability i.e., handling new specifications at runtime without additional training (in the order of several hours).}
\label{tab:comparison}
\scalebox{0.80}{ %
\begin{tabular}{l|ccccccc}
	\hline
	\textbf{Method} & \textbf{\begin{tabular}[c]{@{}c@{}}Het.\\(Specs)\end{tabular}} & \textbf{\begin{tabular}[c]{@{}c@{}}MA\\Aware\end{tabular}} & \textbf{\begin{tabular}[c]{@{}c@{}}Cmplx.\\Specs\end{tabular}} & \textbf{Scale} & \textbf{Div.} & \textbf{Gen.} & \textbf{\begin{tabular}[c]{@{}c@{}}Plan.\\Time\end{tabular}} \\ \hline
	\pwlmastl~\citep{Sun2022}        & $\checkmark$ & $\checkmark$ & $\times$      & $\times$      & $\times$      & $\checkmark$ & \cellcolor{red!20}$>10$\,s \\
	\stlpy~\citep{kurtz2022mixed}             & $\checkmark$ & $\times$     & $\checkmark$  & $\times$      & $\times$      & $\checkmark$ & \cellcolor{red!20}$>10$\,s \\
	\cenlplan~\citep{dawson_robust_2022}             & $\checkmark$ & $\times$     & $\checkmark$  & $\times$      & $\times$      & $\checkmark$ & \cellcolor{red!20}$>10$\,s \\
	\planner~\citep{eappen2024scaling}           & $\times$     & $\checkmark$ & $\checkmark$  & $\checkmark$  & $\times$      & $\times$     & \cellcolor{green!80!black}$<0.1$\,s \\ \hline
	\textbf{\diffsaplanner} (\citep{feng_ltldog_2024})  & $\checkmark$ & $\times$     & $\checkmark$  & $\checkmark$  & $\checkmark$  & $\checkmark$ & \cellcolor{green!60}$\sim1$\,s \\
	\textbf{\diffplanner (Ours)} & $\checkmark$ & $\checkmark$ & $\checkmark$  & $\checkmark$  & $\checkmark$  & $\checkmark$ & \cellcolor{green!60}$\sim1$\,s \\ \hline
\end{tabular}

}
\vspace{-10pt}
\end{table}
We consider Multi-Agent STL in the sense of \oldmastl~\citep{Sun2022}, where a specification $\Psi=\bigwedge_{i=1}^N \phi_i$ constrains each of $N$ agents (e.g., $\phi_i=\E{0}{T}(A_i)\land\E{0}{T}(B_i)\land\E{0}{T}(C_i)$ requires agent $i$ to cover $A_i,B_i,C_i$ within horizon $T$). 
A direct optimization approach that computes a piecewise-linear (PWL) trajectory with $K$ segments per agent~\citep{Sun2022} does not scale with specification complexity and team size, especially once inter-agent interactions are included. 
In particular, pairwise collision-avoidance constraints introduce $\mathcal{O}\!\left(\binom{N}{2}K^{2}\right)$ additional decision variables, which quickly increases problem size.
Recent work \citep[Sec.~3]{eappen2024scaling} provides empirical evidence, reporting timeouts even for simple STL tasks with $N=5$ single-integrator agents in a 2-D space. 
To address this, they adopt an \emph{asynchronous} execution model combined with a GNN-based planner (\planner) and runtime safety filtering via control barrier functions~\citep{zhang_gcbf_2024}, demonstrating results up to $N=32$ agents in a nonlinear setting. 
This execution model differs from \oldmastl, and we refer to the asynchronous variant as \mastl~(formalized in Sec.~\ref{sec:ps-intro}).

An important caveat is that the \planner~planner applies only to homogeneous tasks where all agents follow the same specification. 
This is restrictive, as real-world agents often have diverse objectives, such as covering different goals. Further, it does not allow zero-shot generalization \citep{pmlr-v139-vaezipoor21a} to new specifications since the specification is fixed at training time.
While the approach handles different numbers and positions of agents during testing, a new specification $\phi'$ would require retraining times in the order of several hours.

Moreover, the recurrent planner generates similar plans across agents, causing clustering-induced deadlocks and collisions in practice.
Therefore, ideally, our final approach should possess three general characteristics:

\begin{itemize}
    \item \emph{Scalability}: Can the method efficiently accommodate large numbers of agents?
    \item \emph{Generalizability}: Can the approach handle different goals among agents and new formulas over rectangular predicates placed within the trained goal region at runtime?
    \item \emph{Diversity}: Are generated plans dispersed, thus reducing congestion?
\end{itemize}

Table \ref{tab:comparison} summarizes \mastl~planner capabilities along these dimensions. Inspired by the success of diffusion models in robotics \citep{carvalho2023motion} to model complex objectives and dynamics,
 we propose a novel approach to \emph{multi-agent planning} that addresses these challenges.
Our contributions are as follows: 
1) We propose \diffplanner, a diffusion-based framework for multi-agent planning that generates diverse, jointly-optimized trajectories sampled from a trained single-agent diffusion model.
2) Our approach simultaneously satisfies \mastl~and team-level Capability Temporal Logic-based (\teamspecs\ \citep{liu2023robust}) specifications while incorporating inter-agent collision avoidance.
3) \diffplanner \ enables test-time generalization to new specifications whose predicates are sampled within the trained goal region, without requiring retraining.

\begin{figure}[t]
	\centering
	
	\begin{subfigure}{\linewidth}
		\centering
		\includegraphics[width=\linewidth]{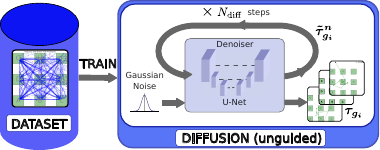}
		\caption{\small Training process for the diffusion model on single agent trajectories (Sec.~\ref{sec:app-traindiff}) without any STL-based guidance.}
		\label{fig:training_diffusion}
	\end{subfigure}

	\begin{subfigure}{\linewidth}
		\centering
		\includegraphics[width=\linewidth]{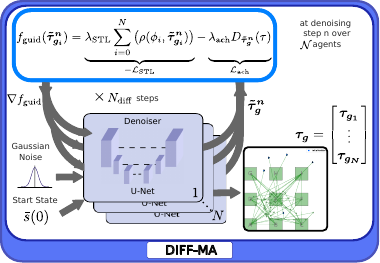}
		\caption{\small Our approach (\diffplanner) showing the use of a single-agent diffusion model for joint STL-guided multi-agent plan generation (Alg. \ref{alg:diffusion-guided} Lines \ref{alg1:lineA}-\ref{alg1:lineC}) for the agents $\agents$ with a given specification $\maspec=\fullmaspec$ (Sec. \ref{sec:app-diff}). $\tautilde[n]{}$ represents denoised goal trajectories for all agents at diffusion step $n$.}
		\label{fig:diffusion_ma}
	\end{subfigure}
	
	\vspace{0.6em} %
	
	\caption{\small \textbf{Top:} Training overview, \textbf{Bottom:} MA STL-guided inference.}
	\label{fig:diffusion_pipeline}
	\vspace{-1em}
\end{figure}

\vspace{-1em}
\subsection*{Related Work}
\label{sec:related-work}
Scalable control in high-dimensional environments often separates temporal logic planning from low-level tracking. High-level approaches include automata and optimization-based STL planners \citep{Icarte2022,liu2025scalable}, as well as multi-agent methods that coordinate or distribute specifications across agents \citep{eappen2024scaling}. Differentiable STL planners instead optimize trajectories through robustness gradients \citep{kapoor_stlcg_2025}. The resulting plans can be tracked using hierarchical RL \citep{Nachum2018} or controllers paired with classical planners \citep{Xiong2022}. Neural Control Barrier Function (CBF) and Graph CBF (GCBF) methods filter actions during execution to preserve safety \citep{qin2021learning,zhang_gcbf_2024}, making them complementary to planners that synthesize trajectories satisfying temporal tasks.

A growing body of work uses diffusion and flow matching models as priors for robot policy and trajectory synthesis \citep{urain2024deepgenerativemodelsrobotics,ho2020denoising,carvalho2023motion}. Temporal logic guidance has also been applied to generative planners, but direct STL guidance often achieves only moderate success in the single-agent setting \citep{kapoor_stlcg_2025,meng2025telograf,liu2025zeroshot}. \citet{shaoul2025multirobot} considers multi-robot planning through a Multi-Agent Path Finding formulation, but assumes perfect tracking to avoid collisions, which is difficult in practice. In contrast, our method adapts the EDM sampler \citep{Karras2022} to guide a single-agent diffusion prior at inference time toward safe joint plans that satisfy STL for many agents, with an emphasis on scalable planning and reliable execution.

\section{Background}

\subsection{Multi Agent Systems with Partial Observability}
\label{sec:bg-mas}
A multi-agent system (MAS) with $N$ agents can be denoted as $\agents=\{1, 2, \ldots, N\}$. 
Each agent $i$ possesses a state $\statevar[t]{i} \in \mathcal{S}_i \subset \mathcal{R}^n$ and can execute an action $\action[t]{i} \in \mathcal{U}_i \subset \mathcal{R}^m$. The evolution of the agents' states is determined by the dynamics function $\statevar[t+1]{i} = f_i(\statevar[t]{i}, \action[t]{i})$. For simplicity, we consider that all agents share the same dynamics function $f_i = f$, identical state space $\mathcal{S}_i = \mathcal{S}$, and action space $\mathcal{U}_i = \mathcal{U}$.
A system trajectory $\tau_s$ is defined as a sequence of states $\tau_s = (\sbar[0], \sbar[1], \ldots, \sbar[T_h])$, where $T_h$ represents the time horizon. 
The state trajectory for an agent $i$ is denoted by $\tau_{s_i} = (\statevar[0]{i}, \statevar[1]{i}, \ldots, \statevar[T_h]{i})$.
Here, $\sbar[t] = (s_1(t), \ldots, s_N(t))$. The system's state is partially observable, meaning each agent can only perceive its own state and the states of other agents within its sensing range \( R \) given as input to its controller.
Each agent follows a policy $\pi_i$, which is a function mapping the agent's state to an action, expressed as $u_i = \pi_i(\sbar_{\mathcal{N}_i})$ where \( \mathcal{N}_i \) is the neighborhood of agent \( i \) within radius \( R \) (Sec. \ref{sec:bg-gcbf}).

\vspace{-0.5em}
\subsection{Signal Temporal Logic}
\label{bg-stl}
Signal Temporal Logic (STL) extends first-order logic with time-bounded variants of the temporal operators from Linear Temporal Logic (LTL). The Boolean connectives are $\land$ (and), $\lnot$ (not), and $\lor$ (or), 
and the temporal operators are $\E{a}{b}$ (eventually within the interval $[a,b]$), $\G{a}{b}$ (always on $[a,b]$), and $\U{a}{b}$ (until within the interval from $[a,b]$). The syntax is as follows:
	\begin{equation*}
\phi := \mathcal{P} \mid \lnot \phi \mid \phi \land \psi \mid \phi \lor \psi %
		 \mid \E{a}{b} \phi \mid \G{a}{b} \phi \mid \phi \U{a}{b} \psi
		 \end{equation*}

where $\mathcal{P}$ denotes a predicate that maps states to real values. The quantitative (robustness) semantics of STL \citep{maler2004monitoring} assign a real value $\rho(\phi,\tau_{s_i})$ indicating the degree to which a state trace $\tau_{s_i}$ satisfies or violates $\phi$. This robustness can be differentiable \citep{LeungArechigaEtAl2021}, which enables direct gradient-based optimization of STL specifications using differentiable planners, including neural network-based methods.

In the context of multi-agent systems with $N$ agents, the \mastl~specification $\Psi$ is constructed from $N$ separate STL specifications, represented as $\bigwedge_{i=1}^N \phi_i$, where $\phi_i$ corresponds to the STL specification for each individual agent indexed by $i$. The \mastl~specification $\Psi$ is fulfilled when all individual STL specifications are met and there are no collisions among the agents.
Similar to the variant of STL depicted in \citet{Sun2022}, \mastl~is a subset of all STL specifications defined on the joint state space, yet captures many intended multi-agent behaviors.

We also consider team-level specifications in the form of \teamspecs~\citep{liu2023robust}, whose atomic unit is a \emph{task} $\ctask = \catltask{\phi}{\taskcount}$\footnote{In this paper, we consider identical agents sharing a single capability~\citep{liu2023robust}, so the capability argument is omitted from the original task tuple.} satisfied when at least $\taskcount$ agents meet an inner-logic formula $\phi$ on their individual trajectories.
The outer logic composes tasks via Boolean connectives, e.g.\ $\ctask_1 \land \ctask_2$ requires both tasks to be satisfied while $\ctask_A \lor \ctask_B$ requires either task to be satisfied.
Using \mastl\ as the inner logic, this framework covers both per-agent independence ($\taskcount\!=\!1$) and inter-dependent team tasks ($\taskcount\!>\!1$) where multiple agents must coordinate on a shared formula.

\vspace{-0.5em}
\subsection{Graph-Based Modeling and Safety via Graph Control Barrier Functions}
\label{sec:bg-gcbf}

Our controller (\gcbfp~\citep{zhang_gcbf_2024}), uses a \emph{Graph Control Barrier Function} (\gcbf), and adopts a standard graph-based abstraction for multi-agent systems (MAS), where agents are nodes of a graph \( \graph = (\vertices, \edges) \). Edges encode local interactions with an edge \( (\vertice{i}, \vertice{j}) \in \edges \) being present when the Euclidean distance between nodes \( \vertice{i} \) and \( \vertice{j} \) does not exceed a sensing radius \( R \). %
In the \gcbfp~controller, a graph neural network (GNN) processes this representation and parameterizes a local control policy for each agent.
\gcbfp~supports both agent-agent and agent-obstacle interactions, but we focus on the former for our scalability analysis due to a known limitation of deadlocks when obstacles are present \cite{zhang2025dgppo}.
The safe set \( \safeset \subset \mathcal{S}^N \) for an \( N \)-agent MAS (where $\mathcal{S}^N$ denotes the joint state space over $N$ agents) is the collection of states \( \sbar \) satisfying the inter-agent separation requirements in Definition ~\ref{def:ps-STL-MA}:
\begin{equation} \label{eq:SN_def}
 \begin{aligned}
    \safeset \coloneqq \Big\{ \sbar \in \mathcal{S}^N \;\Big|
    \;  \min_{i, j \in \agents, i \neq j} \norm{p_i - p_j} > 2r \Big\}
 \end{aligned}
\end{equation}
where \( r \) is the radius of each agent, and \( p_i \) is the position of agent \( i \).
The complement \( \unsafeset \coloneqq \mathcal{S}^N \setminus \safeset \) is the unsafe set. A continuously differentiable function \( h : \mathcal{S}^M \to \mathbb{R} \) is a \gcbf~if there exists an extended class-\( \mathcal{K}_\infty \) function \( \alpha \) and a local policy \( \pi_i : \mathcal{S}^M \to \mathcal{U} \) for each agent \( i \in \vertices_a \) such that, for all \( \sbar \in \mathcal{S}^N \) with \( N \ge M \),
\begin{equation}
\label{eq:graph CBF}
\dot h(\sbar_{\mathcal{N}_i}) + \alpha\!\big( h(\sbar_{\mathcal{N}_i}) \big) \ge 0, \quad \forall i \in \vertices_a,
\end{equation}
\begin{equation}
\label{eq:hdot_def}
\dot h(\sbar_{\mathcal{N}_i}) = \sum_{j \in \mathcal{N}_i}
\frac{\partial h(\sbar_{\mathcal{N}_i})}{\partial \statevar{j}}\, f^c(\statevar{j}, \action{j}).
\end{equation}
Here $f^c$ denotes the continuous-time agent dynamics, whose sampled-data discretization yields the discrete-time model in Sec.~\ref{sec:bg-mas}, and the controller enforces \eqref{eq:graph CBF} at each sampling instant.
Under these conditions, the zero-superlevel set of $h$ is forward invariant: if a policy \( \pi_i \) and a \gcbf\ \( h \) satisfy \eqref{eq:graph CBF} for all agents and all \( \sbar \in \safeset \), then trajectories starting in the zero-superlevel set of $h$ (contained in $\safeset$) do not enter \( \unsafeset \) \citep{zhang_gcbf_2024}. 

\vspace{-0.5em}
\subsection{Diffusion Models as Trajectory Generators}
\label{sec:diffusion}
Diffusion models~\citep{ho2020denoising, Karras2022} are a class of generative models that have recently shown great promise in modeling complex data distributions, particularly in high-dimensional spaces. They are designed to model the way data can be transformed through a diffusion process, which progressively adds noise to data and then learns to reverse this process to generate new samples. In the context of planning, diffusion models can be used to generate state trajectories~\citep{Janner2022} that adhere to system dynamics and satisfy certain constraints or objectives. Our framework follows techniques in \citet{Karras2022} (EDM) where, for a data point $\bm{x}$, each denoising step $i$ of $\ndiff$ steps of an Ordinary Differential Equation (ODE) for a noise schedule $\sigma(i)$ applies $d\bm{x} = -\dot{\sigma}(i)\sigma(i)\nabla_{\bm{x}} \log p\left(\bm{x}; \sigma(i) \right) di$,
where \mbox{$\dot{\sigma} = \frac{\mathrm{d}\sigma}{\mathrm{d}i}$} and $\nabla_{\bm{x}} \log p\left(\bm{x}; \sigma(i)\right)$ is the score function.  
Following EDM, we estimate the score from a learned denoiser 
$D_\theta(\bm{x};\sigma)$ with weights $\theta$.
As a convention, throughout this work we will highlight the noised (marked $\boldsymbol{\hat{x}}$) and denoised outputs (marked $\boldsymbol{\tilde{x}}$) of diffusion models in bold.

\vspace{-0.5em}
\section{Problem Statement}
\label{sec:stl-mamp}
\label{sec:ps-intro}

Consider $\Psi$ to be either an \mastl\ specification
  ($\Psi = \bigwedge_{i=1}^N \phi_i$) or a \teamspecs\ team specification
  (Sec.~\ref{bg-stl}) for $N$ agents $\agents=\{1,\dots,N\}$ with positions $\positionvar[t]{i}\in\positionspace\subset\R^n$ ($n\in\{2,3\}$). Each agent's state $\statevar[t]{i}\in\statespace$ maps to its position via the projection $\filterstate:\statespace\rightarrow\positionspace$, i.e., the first $n$ components of $\statevar[t]{i}$. Agents are modeled as balls of radius $r>0$: when an agent is at $p\in\positionspace$, it occupies $B_r(p)$. Given local controllers $(\pi_i)_{i\in\agents}$, we aim to design a centralized planner $\jointplannn$ that produces per-agent goal sequences $\goaltraji{i}=(\goal[0]{i},\ldots,\goal[T]{i})$ of length $T<T_h$ over a planning horizon $T_h$, while execution is decentralized under $(\pi_i)_{i\in\agents}$.

\begin{definition}[Motion planning problem under \mastl/\teamspecs]\label{def:ps-STL-MA}
Given $\Psi$, agents $\agents$, and policies $(\pi_i)_{i\in\agents}$ as above, a solution is a planner $\jointplannn$ such that the resulting closed-loop trajectories over $[0,T_h]$ satisfy:
\begin{itemize}
    \item \textbf{Agent safety.} For all $t \in [0, T_h]$, and for all $i, j \in \agents$ where $i \neq j$, $\norm{\positionvar[t]{i} - \positionvar[t]{j}} > 2r$
    \item \textbf{Asynchronous STL satisfaction.} For each agent $i\in \agents$ there exists $t_{i,0}<t_{i,1}<\dots<t_{i,T}$  such that $t_{i,k} \in \left(0, \ldots, T_h\right)$ and the trajectories $\tau_i = (\statevar[t_{i,0}]{i}, \dots, \statevar[t_{i,T}]{i})$
    jointly satisfy $\Psi$
  under the robustness semantics of Sec.~\ref{bg-stl},
  i.e.\ $\rho(\Psi, \{\tau_i\}_{i\in\agents})\geq 0$.
  For \mastl\ this reduces to $\rho(\phi_i,\tau_i)\geq 0$ for all $i$.
    \item \textbf{Achievability of goals.} For all $i\in \agents$, given the goal trajectory $\traj_{\goal{i}}$  of length $T$ from $\plannn{i}$, the tracking deviation
 $
    \trajdist{\traj_{\goal{i}}}{\traj_i}= \sum_{t'=0}^{T} \norm{\filterstate(\statevar[t_{i,t'}]{i}) - \filterstate(\goal[t']{i})}_2 < \epsilon
 $  for a small %
 $\epsilon \in \R^+$. %
\end{itemize}
\end{definition}
\noindent

\vspace{-1em}
\section{Approach}
\label{sec:app-intro}

We implement a diffusion guidance approach 
which utilizes the diffusion process (EDM) in \citet{Karras2022} to guide the plans generated by the diffusion model towards satisfying the STL objectives 
while respecting multi-agent safety constraints. 
Our solution, \diffplanner~(shown in Fig.~\ref{fig:diffusion_ma}), is a two-stage process that first trains a diffusion model on single-agent trajectories (Sec. \ref{sec:app-traindiff}, Fig.~\ref{fig:training_diffusion}) covering a set of STL predicates $\Predset$, 
and then uses the trained model (Sec. \ref{sec:app-diff}) to jointly generate multi-agent trajectories ($\bartautilde$) for a system of agents (Sec. \ref{sec:app-multiagent}) guided by a new STL specification at test time. 
Training remains single-agent because the multi-agent coupling is injected at test time through the STL and achievability guidance rather than learned in the prior.
We embed \gcbfp\ (Sec.~\ref{sec:bg-gcbf}) in guidance (via closed-loop rollouts to calculate the achievability loss $\lossacheivable$) and execution, promoting compliance with safety requirements.

This process allows for safe, test-time generalization to new STL objectives involving arbitrary compositions of rectangular predicates placed within the goal region covered during training, following rules described in Sec. \ref{bg-stl}.

\vspace{-1em}
\subsection{Training Diffusion Models for Downstream STL Tasks}
\label{sec:app-traindiff}

The first stage involves fixing a set of Predicates $\Predset$ which are used in our STL objectives.
For our work, we assume rectangular predicate regions as shown in Fig.~\ref{fig:robot_demo}.%
With this set of predicates, we sample trajectories for a single agent achieving various randomly sampled sequences of goals following our same controller (\gcbfp) to get a dataset covering the state space or a set of trajectories $\tau_i$ for a single agent $i$ with a time horizon $T_h$.
Since our objective is to obtain a planner $\jointplannn$ to generate $\goaltraj$ for the multi-agent system, we fix the planning length $T$ and sample the states $\statevar[t]{i}$ and goals $\goal[t]{i}$
for each agent $i$ at each time step $t=\frac{T_h}{T}j$ for $j=0,1,\ldots,T$. 
Given this dataset, we run a supervised training process where we train a diffusion model (Fig.~\ref{fig:training_diffusion}) to generate state-goal trajectories that represent $\btau_{i}=(\taustate[]{i}, \taudiff[]{i})$  for an agent $i$.

\begingroup
\makeatletter
\let\RAL@fs@ruled\fs@ruled
\def\fs@ruled{%
	\RAL@fs@ruled
	\let\RAL@old@fs@pre\@fs@pre
	\def\@fs@pre{\vspace*{5pt}\RAL@old@fs@pre}%
}
\makeatother
\begin{algorithm}[t]
	
	\footnotesize      
	\flushleft
	\caption{Plan sampling via {\color{pgd}STL-guided} diffusion
			\citep{Karras2022}.}
		\label{alg:diffusion-guided}
		
			\begin{algorithmic}[1]
				\State {\bfseries Parameters:} Noise levels $\sigma^n$, guidance levels $\lambda^n$, noise factor $\gamma^n$, noise scale $S_{\text{noise}}$, diffusion steps $\ndiff$, small constant $\epsilon_g$
				\State {\bfseries Required:} Denoiser model $D_{\theta}$, guidance function $\guidancefn$,
				\State {\bfseries sample} $\btau[0] \sim \mathcal{N}(\bm{0}, \sigma_0^2\bm{I})$ \AlignedComment{Sample random noise trajectory}
				\For{$n=0$ {\bfseries to} $\ndiff-1$}
				\State {\bfseries sample} $\epsilon^n \sim \mathcal{N}(\bm{0}, S^2_{\text{noise}}\bm{I})$ \hspace{3mm}\Comment{Temporarily raise noise}
				\State $\hat{\sigma}^n \leftarrow \sigma^n + \gamma^n \sigma^n$
				\State $\btauhat[n] \leftarrow \btau[n] + \sqrt{(\hat{\sigma}^n)^2 - (\sigma^n)^2}\bm{\epsilon}^n$
				\State $\btautilde[n] \leftarrow \bigl[D_{\theta}(\btauhat[n]_{i};\hat{\sigma}^n)\bigr]_{i=1}^N$ \hspace{10mm} \Comment{Estimate denoised trajectory}
				\State $\bm{d}^n \leftarrow \left(\btauhat[n] - \btautilde[n]\right)/\hat{\sigma}^n$ \hspace{20mm} \Comment{Evaluate $\frac{\partial \bm{\tau}}{\partial \sigma}$ at $\hat{\sigma}^n$} \color{pgd}
				
				\State $\bm{g}^n \leftarrow \nabla_{\tautilde[n]{}} \guidancefn(\tautilde[n]{})$ \hspace{8.5mm} \label{lst:guidancestep} \Comment{Denoised plan gradient} \label{alg1:lineA}
				\State $\lVert\bm g^n\rVert_{2,\text{agent}}
				\triangleq \bigl[\lVert \bm g^{n}_{1}\rVert_2,\dots,
				\lVert \bm g^{n}_{N}\rVert_2\bigr]^\top$
				\Comment{\footnotesize Agent‑wise norm} \label{alg1:lineB}
				\State $\tauhat[n]{}\leftarrow\tauhat[n]{}
				+\lambda^n\!\left(\bm g^n\oslash(\lVert\bm g^n\rVert_{2,\text{agent}}+\epsilon_g)\right)$ \label{alg1:lineC}
				\Statex \Comment{\mastl~guidance ($\oslash$ : agent-axis division,  $\epsilon_g > 0$ for stability)} \hspace{-1mm}\color{black}
				\State $\btau[n+1] \leftarrow \btauhat[n] + (\sigma^{n+1} - \hat{\sigma}^n)\bm{d}^n$ \hspace{13mm}\Comment{Apply Euler step}
				\If{$\sigma^{n+1} \neq 0$}
				\State $\bm{d}'^n \leftarrow \left(\btau[n+1]  - \bigl[D_{\theta}(\btau[n+1]_i ;\sigma^{n+1})\bigr]_{i=1}^N\right)/\sigma^{n+1}$ 
				\Statex \Comment {2nd order correction over $N$ agents}
				\State $\btau[n+1] \leftarrow \btauhat[n]  + (\sigma^{n+1} - \hat{\sigma}^n)\left(\frac{1}{2}\bm{d}^n + \frac{1}{2}\bm{d}'^n\right)$
				\EndIf
				\EndFor
				\State {\bfseries return} $\taudiff[\ndiff]{}$
			\end{algorithmic}

	\end{algorithm}
\endgroup

\subsection{Guiding Diffusion Models with STL Objectives}
\label{sec:app-diff}

The \diffplanner~diffusion framework, presented in Algorithm~\ref{alg:diffusion-guided}, is based on the idea of using the denoised trajectory at a given diffusion step\footnote{We use superscripts on variables to show diffusion timesteps and subscripts to denote agent indices.} $n$
to estimate the gradient of the trajectory with respect to the STL robustness score $\rho$ (\sectionautorefname~\ref{bg-stl}).
This gradient is then used to guide the trajectory towards satisfying the STL objectives.
The initial condition is imposed by inpainting, overwriting the first plan step with the agent's current state after each denoising step \citep{Janner2022}.

We use $\btau[n]$ to denote the state-goal trajectory at diffusion step $n$ and $\taustate[n]{}$ and $\taudiff[n]{}$ to denote the state and goal components of the trajectory, respectively.
Our guidance function ($\guidancefn$) combines both the STL robustness score and a notion of achievability to steer the trajectory towards satisfying the STL objectives while ensuring that the generated trajectory is achievable by the agents in the system while following the \gcbfp~controller.
To this end we define the guidance function as follows, where $\rho_{\phi}$ is the STL robustness score and $\lossacheivable$ is the achievability loss function $\guidancefn = - \coeffstl\lossstl - \coeffacheivable\lossacheivable$
where $\coeffstl$ and $\coeffacheivable$ are positive weights for the STL loss and achievability loss, respectively.
The STL loss ${\lossstl}_i$ for agent $i$ is defined as the negative of the STL robustness score $\rho(\phi_i, \tautilde[n]{i})$,
with the total $\lossstl=\sum_{i=1}^N {\lossstl}_i$ for all agents $\agents$.
 For \teamspecs~specifications, whose robustness is not agent-decomposable \citep{liu2023robust}, we instead set $\lossstl = -\rho(\Psi, \taudiff[n]{})$ using the joint multi-agent trajectory.
The achievability loss function $\lossacheivable$ follows ideas from prior work \cite{eappen2024scaling} and is the tracking error between the denoised trajectory $\bartautilde[n]$ and a sampled trajectory using the \gcbfp~controller.
We overload the notation of trajectory distance for the system of agents $\lossacheivable = \trajdist{\bartautilde[n]}{\btau[n]} = \sum_{i=1}^N \trajdist{\tautilde[n]{i}}{\taustate[n]{i}}$.
By minimizing this loss, we ensure that the generated trajectory is cognizant of the system dynamics and the controller's capabilities in the presence of other agents.
Since calculating $\trajdist{}{\btau[n]}$ is expensive, we find it sufficient to include this in the last fraction ($\nachfrac$, e.g. 20\%) of diffusion steps avoiding wasted computation on the initial near-Gaussian samples.
After $\ndiff$ steps the final plan $\taudiff[\ndiff]{}$ is generated guided by both the MA-STL objective and our multi-agent achievability objective.

\subsection{Applying Diffusion Planning to Multi-Agent Systems}
\label{sec:app-multiagent}

Diffusion inference is inherently slow \citep{li2024faster}, and multi-agent joint sampling compounds this cost. We use a JAX-based \citep{jax2018github} JIT-compiled vectorization of Algorithm~\ref{alg:diffusion-guided} for fast inference.
However, diffusion samples may still violate STL constraints. Related work \citep{feng_ltldog_2024,kapoor_stlcg_2025,meng2025telograf} reports only 50--80\% success, which worsens in multi-agent settings.
Our sampler therefore \emph{resamples} each agent $i$'s plan before execution, up to $\maxresample$ times, until $\rho(\phi_i,\taudiff[]{i})>\epsresample>0$ (included in reported planning time).
Each iteration draws $B=8$ candidate joint plans in one reverse pass and keeps the one with highest exact robustness (joint team robustness for \teamspecs), never mixing agents across candidates.
Accepted plans are frozen and excluded from further updates, so the accepted set is monotonically non-decreasing (for \teamspecs, acceptance uses the joint test $\rho(\Psi,\taudiff[]{})>\epsresample$ without per-agent freezing).
A resampled agent keeps its previous plan unless a new draw scores strictly higher exact robustness, so kept robustness is monotone non-decreasing, and at the cap ($\maxresample$) the loop returns the highest-robustness plan found.
Acceptance certifies STL satisfaction at the plan level only. The achievability condition in Definition~\ref{def:ps-STL-MA} is promoted through the soft guidance term $\lossacheivable$ rather than formally guaranteed, and is assessed empirically through the reported success rates.

\section{Experiment Setup}
All experiments are analyzed with respect to the three characteristics
introduced in Section~\ref{sec:intro}:

\centerline{\textbf{C1}: Scalability \qquad
            \textbf{C2}: Generalizability \qquad
            \textbf{C3}: Diversity.}

\subsection{Performance at Scale (C1)}

We validate our method on the DubinsCar benchmark, a widely used non-linear testbed for multi-agent navigation~\cite{zhang_gcbf_2024}. 
Our framework is constructed using JAX \citep{jax2018github}, building on the diffusion model framework provided by \citet{jackson2024policy} and \gcbfp~ \citep{zhang_gcbf_2024} (see Sec.~\ref{sec:bg-gcbf}), with \emph{all} comparisons employing this underlying collision avoidance controller.
We also include a video demonstration on a real differential-drive robot task run via the Robotarium test-bench \cite{CSS:robotarium} 
and results on other linear environments 
on our \href{\websiteurl}{website}. 
All diffusion training and inference procedures were run on an AWS \verb|g6.2xlarge| instance or equivalent with 8 Intel Xeon-based CPU Cores and 32 GB of RAM with an Nvidia L4 GPU.
We use  256 diffusion steps $(\ndiff)$ and up to  40 plan resamples $(\maxresample)$ with the noise levels $(\sigmax,\sigmin)$ following the default settings of \citet{jackson2024policy}.

We vary the number of agents and benchmark against the strongest available baselines on various tasks.
Since no existing multi-agent STL planner supports this setting at the scales considered, the fairest comparison is to execute state-of-the-art single-agent planners independently for each agent: a MILP-based planner (\emph{\stlpy})~\citep{kurtz2022mixed}, a counterexample-guided gradient method (\emph{\cenlplan})~\citep{dawson_robust_2022}, and a single-agent ablation of our approach (\emph{\diffsaplanner}), which can be viewed as a multi-agent extension of \citet{feng_ltldog_2024} with STL-score--based resampling. 
These baselines do not model inter-agent interactions during planning (Table~\ref{tab:comparison}) and therefore rely on \gcbfp{} for collision avoidance (Sec.~\ref{sec:bg-gcbf}) at execution time. 
Separately, to analyze plan diversity (Table~\ref{tab:results-diversity}) in the \emph{homogeneous} setting (where all agents share the same specification), we include the \emph{\planner} planner~\citep{eappen2024scaling}, which does not support heterogeneous tasks.
Collision-aware MILP approaches are known not to scale past five agents (\(N>5\))~\cite[Sec.~3]{eappen2024scaling}, making this the most sensible comparison.

\begin{figure}[t!]
    \centering
    \includegraphics[width=\columnwidth]{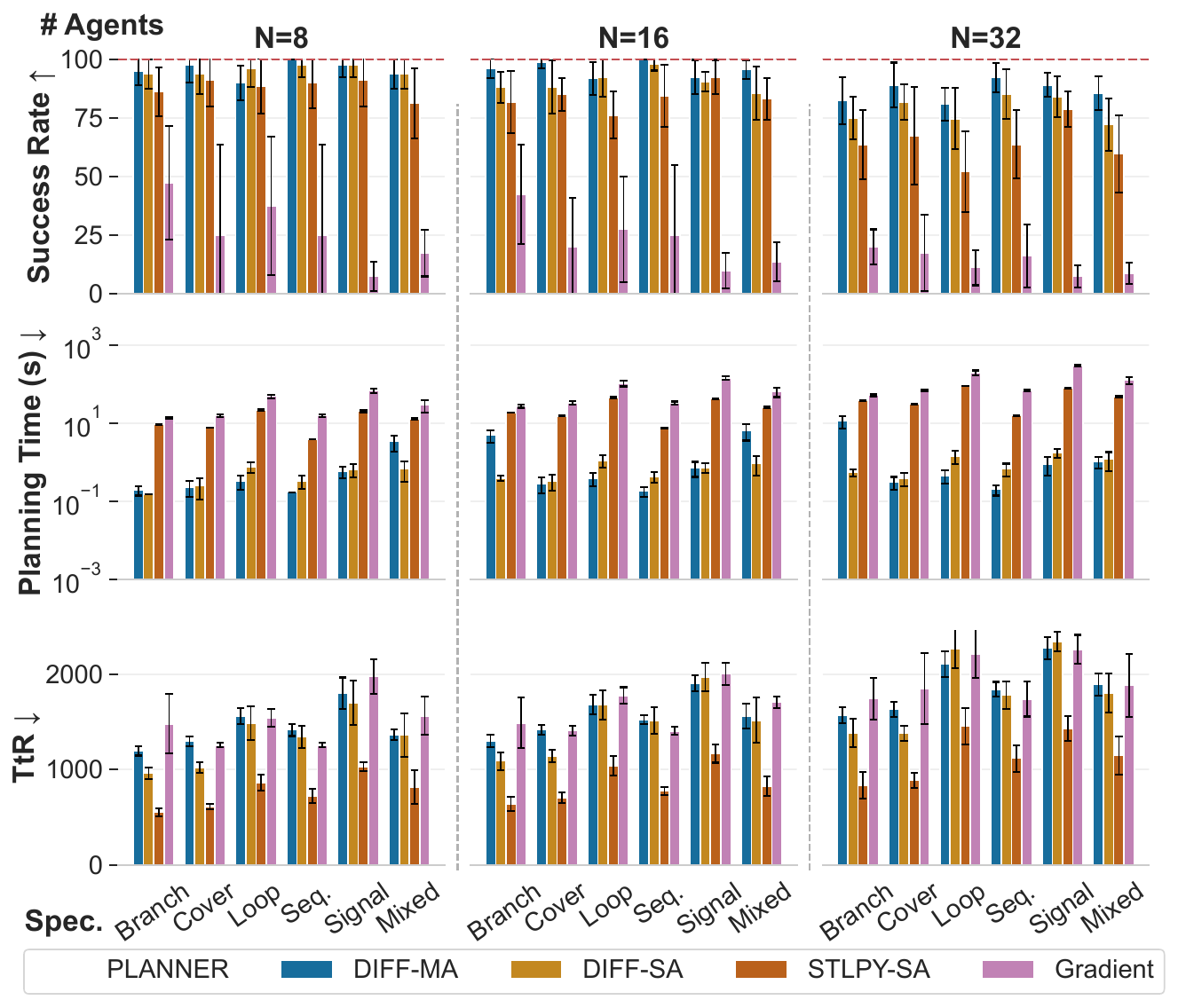}
    \caption{\small
Results over different number of agents and specifications.
		\textbf{Top:} Success Rate (\%).
		\textbf{Middle:} Planning Time (s).
		\textbf{Bottom:} Time-to-Reach (TtR, steps).
		Standard deviations are shown around the mean.
        Our method (\diffplanner)~on average exceeds  \stlpy~by 20\% with a notable relative success rate improvement of 36\% for the crowded map ($N=32$), while being 55x faster with zero retraining.
	}
	\label{fig:all-results-combinedDubins}
    \vspace{-1em}
\end{figure}

We report the mean \emph{Planning Time} (in seconds), 
the \emph{Success Rate} (percentage of runs where the STL specification was satisfied and no collisions occurred), 
and the \emph{TtR} (time-to-reach, the steps successful runs take to complete the task) over 30 seeds (see Fig. \ref{fig:all-results-combinedDubins}).
The effect of the achievable-loss coefficient $\coeffacheivable$ is studied in an ablation with $\coeffacheivable{=}10^{-3}$ (LA) vs.\ $\coeffacheivable{=}1$ (full) at fixed $\coeffstl{=}1$ (Table~\ref{tab:mixed-extract-compact-na}).
We further compare (Fig.~\ref{fig:delay-per-agent}) the per-agent execution delay of \diffplanner~due to runtime maneuvering against the time an isolated single agent (planned with \stlpy) takes to satisfy its task on the \mixedspec\ specification.
Lastly, we evaluate \diffplanner\ on a larger map ($2\times$ the grid size) with up to $N=128$ agents for the \mixedspec\ specification to study scalability at high $N$ (Fig. \ref{fig:highN-results}).

\vspace{-0.5em}
\subsection{Generalizing to Different Goals (C2)}
\label{sec:exp-setup-specs}
To get a dataset for training our diffusion model, we first define a set of predicates $\Predset$ that we want to cover in our dataset.
Our predicates are defined as rectangular regions in the 2D space, and placed evenly in a grid of $4.0 \times 4.0$ units.
We use 9 predicates in total, which are shown in green in Fig. \ref{fig:robot_demo}.
To create a dataset to train our diffusion model that covers our predicate space $\Predset$, %
we sampled 10000 trajectories using the \stlpy~planner on a \seq~specification with 3 predicates randomly sampled from $\Predset$.
During inference (for the \mastl\ results in Fig. \ref{fig:all-results-combinedDubins}, \ref{fig:delay-per-agent} ), we sample arbitrary predicate centers from the 4.0 $\times$ 4.0 unit space, with a size sampled uniformly from $[0.5, 1.0]$ units, 
to demonstrate that our diffusion model generalizes to unseen predicates (within the training space).
For the \teamspecs\ (Fig. \ref{fig:team-spec-results}) and diversity experiments (Table \ref{tab:results-diversity}), we use fixed predicates from $\Predset$ on all settings for ease of visualization and comparison.

Unlike the homogeneous setting (Table \ref{tab:comparison},~\citep{eappen2024scaling}), the heterogeneous setting allows for different specifications among agents. 
We evaluate five STL task templates that capture common multi-agent missions --
\emph{sequence} (\seq): $\tWedge_{i=1}^{3}\E{\tau_{i-1}}{\tau_i}(p_i)$ with $p=(X,Y,Z)$ and $\tau=(0,\tfrac{T}{3},\tfrac{2T}{3},T)$, %
\emph{coverage} (\cover): $\tWedge_{p\in\{X,Y,Z\}}\E{0}{T}(p)$, 
\emph{looping over goals} (\loopspec): $\G{0}{T/2}\bigl(\tWedge_{p\in\{X,Y,Z\}}\E{0}{T/2}(p)\bigr)$, 
\emph{branching} (\branch): $\tVee_{S\in\{\{X,Y\},\{Z,W\}\}}\tWedge_{p\in S}\E{0}{T}(p)$, 
\emph{loop until twice} (\signalspec): $\loopspec\,\U{T_\ell}{T}\,W$ (loop over $\{X,Y,Z\}$ twice, then reach a new goal $W$ in the final steps $[T_\ell,T]$, where $T_\ell$ is the loop horizon),
and a \emph{mixed} structure (\Mixedspec): $\phi_i \sim \mathrm{Uniform}(\seq,\cover,\loopspec,\branch,\signalspec)$,
where $T=15$ is the planning horizon, $k=20$ the goal-sampling interval and unique predicates $X,Y,Z,W$ are uniformly sampled for each agent $i$ from the set $\Predset$. %
These encode ordered waypoint visits, unordered area coverage, periodic patrols, spatially partitioned goals, and uniformly sampled heterogeneous missions among agents.
The above specifications assign one formula per agent ($\taskcount = 1$).

We also evaluate two \teamspecs\ tasks (Sec.~\ref{bg-stl}) with redundancy ($\taskcount\!>\!1$) and inter-task avoidance (agents in one task avoid the goals of others):
\teamchoice: $(\ctask^A_0 \land \ctask^A_1) \lor (\ctask^B_0 \land \ctask^B_1)$, a disjunction of two branches, each pairing a 3-stage \seq\ ($\taskcount\!=\!3N/4$: gathering point then two waypoints along one edge) with a support task ($\taskcount\!=\!N/4$: two sequential goals).
\teamredundant: $\tWedge_{q=1}^{4}\ctask_q$, i.e.\ $4$ goals from $\Predset$ each reached by $\taskcount\!=\!N/4$ agents where $\ctask_q = \catltask{\E{0}{T}(p_q)}{\taskcount}$.
With $\taskcount$ agents converging on the same region, independent planning cannot anticipate collision-avoidance maneuvers.
We additionally report the \emph{Task Rate}, the fraction of episodes where the team-level specification is satisfied (Fig.~\ref{fig:team-spec-results}).
Task Rate ignores collisions, whereas Success Rate requires both per-agent task satisfaction and no collisions.
Our baselines here are \stlpyglobal, a global MILP jointly solving allocation (tractable only at $N\!=\!8$, timing out $>10$\,min otherwise), and per-agent \stlpy\ under oracle task-to-agent allocation (best of closest first-goal, random, and Hungarian travel-distance matching). \diffplanner\ requires no allocation scheme.

\vspace{-.5em}
\subsection{Diversity in Generated Plans (C3)}
\label{subsec:diversity}
We evaluate how much the agents \quotes{share} their trajectories by comparing against the \planner~and \stlpy~in the homogeneous setting (Table~\ref{tab:results-diversity}), where each agent had the same specification, using two complementary measures:
(i)~\emph{Path Overlap} uses a 2D occupancy grid of the environment to compute the fraction of cells visited by all agents divided by the cells visited by any agent.
A value of 0 means the plans are completely disjoint, while a value of 1 means all agents have the same trajectory.
(ii)~\emph{Agents per Cluster}: from the pair-wise discrete Fréchet distance matrix \cite{eiter1994computing} we connect two paths when their distance is at most $\dtau=1$ 
and count the average number of agents per connected component.

\begin{table}
	\centering
	
	\caption{\small Comparison of plan diversity using the average number of agents per cluster and path overlap (\%) for different planners (D-MA: \diffplanner, GO: \planner, STLPY: \stlpy) in the Dubins car environment for Homogeneous specifications (same across agents). 
		The best score (w.r.t. diversity) is highlighted in bold. 
	}
	
	\scalebox{0.95}{
		\begin{tabular}{l|l|ccc|ccc}
			\toprule
			&  & \multicolumn{3}{c|}{Agents per Cluster $\downarrow$} & \multicolumn{3}{c}{Path Overlap (\%) $\downarrow$} \\
			Spec. & N & D-MA & GO & STLPY & D-MA & GO & STLPY  \\
			\midrule
			\Branch & 16 & \textbf{1.09} & 3.81 & 2.35 & \textbf{0.00} & 0.08 & \textbf{0.00} \\
			\cmidrule{1-8}
			\Cover & 16 & \textbf{1.08} & 6.40 & 3.02 & \textbf{0.00} & 0.18 & \textbf{0.00} \\
			\cmidrule{1-8}
			\Loopspec & 16 & \textbf{1.04} & 4.40 & 3.40 & \textbf{0.00} & 8.72 & 0.27 \\
			\cmidrule{1-8}
			\Seq & 16 & \textbf{1.08} & 7.74 & 2.89 & \textbf{0.00} & 0.25 & 12.42 \\
			\cmidrule{1-8}
			\bottomrule
		\end{tabular}
	}
	
	\vspace{-1em}
	\label{tab:results-diversity}
	
	\end{table}

\vspace{-0.5em}
\section{Results}
\label{sec:results}
\vspace{-0.5em}

\begin{figure}[t!]
    \centering
    \includegraphics[width=\columnwidth]{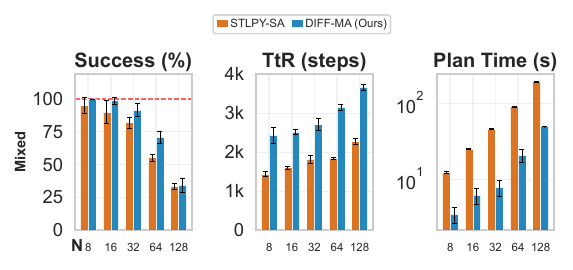}
    \caption{\small
		We evaluate \diffplanner\ on a larger map with up to $N=128$ agents for the \mixedspec\ spec. and note degradation in success at high $N$ (matching \stlpy\ at $N=128$) with modest increases in planning time.
    }
    \label{fig:highN-results}
    \vspace{-1em}
\end{figure}

\begin{figure}[t!]
    \centering
    \includegraphics[width=\columnwidth]{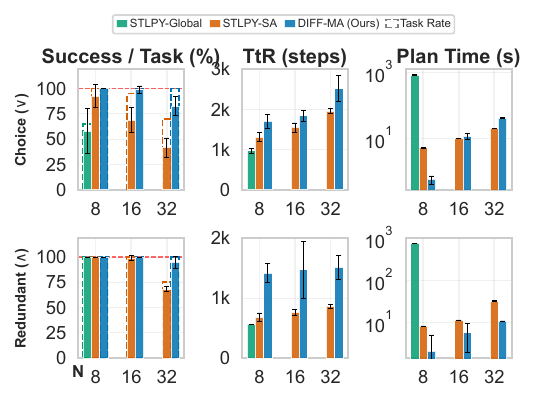}
    \caption{\small
		Team-level spec. results on \teamspecs\ tasks with redundancy ($\taskcount > 1$) and inter-task avoidance constraints:
        \emph{Choice-Seq} (two branches, each a pair of sequential tasks) and \emph{Redundant} (shared waypoint sequences).
        Baselines: \stlpyglobal\ (global MILP over all agents, feasible only at $N\!=\!8$) and \stlpy\ with oracle allocation (best of greedy and Hungarian matching by travel distance).
        \diffplanner\ requires no task-to-agent allocation and maintains high success at scale ($N\!=\!32$) where \stlpy\ degrades sharply.
    }
    \label{fig:team-spec-results}
    \vspace{-1em}
\end{figure}

\noindent\textbf{Performance (C1, C2)} 
Unlike \planner, which cannot accommodate heterogeneous specifications, \diffplanner\ does so and, on average, attains a $20\%$ higher success rate than \stlpy\ while planning $3\times$--$200\times$ faster (Fig.~\ref{fig:all-results-combinedDubins}). 
At the largest evaluated scale ($N\!=\!32$), \diffplanner\ achieves a $36\%$ relative increase in success over \stlpy. 
As with \planner, \diffplanner\ deliberately trades time-to-reach (TtR) for safety: its paths are $\approx 48\%$ longer than those of \stlpy, whose average success at this scale is only $64\%$, to maintain safety in dense multi-agent settings. 
By contrast, \cenlplan\ underperforms across most specifications (never exceeding $20\%$ success on the 32-agent tasks) and shows only marginal improvements on \seq/\cover, while \diffsaplanner\ performs well on smaller instances but falls below $79\%$ average success at this scale. 
Overall, \diffplanner\ remains consistently strong, achieving at least $84\%$ success even with 32 agents. 
We attribute the weaker performance of \cenlplan, \stlpy, and \diffsaplanner\ to their design choices: they optimize individual-agent objectives without modeling inter-agent collisions during planning, instead relying on runtime collision avoidance.
A failure-mode breakdown on our \href{\websiteurl}{website} supports this, as the remaining failures at $N{=}32$ are dominated by collisions, with the collision share of \diffplanner\ (10.9\%) staying $2.5$–$3\times$ below \stlpy\ (34.4\%) and \diffsaplanner\ (27.5\%) while planner-side STL failures remain minimal.

On increasingly dense multi-agent scenarios (Fig.~\ref{fig:highN-results}), the \gcbfp~controller's limited collision avoidance leads to degraded success at high $N$ (matching \stlpy\ at $N=128$), noting that \citet{zhang_gcbf_2024} use a much larger map (roughly $3\times$ our size) for a reasonable safety rate at $N=128$.
This is reflected by the measured safety rate of $42\%$ at $N=128$ for \diffplanner\ (vs.\ $95\%$ at $N=32$).
At that map scale, the diffusion model needed further optimization to cover the larger predicate space, which we leave for future studies.

\begin{figure}[t!]
     \centering
     \begin{subfigure}[b]{\columnwidth}
         \centering
     \includegraphics[width=0.9\textwidth]{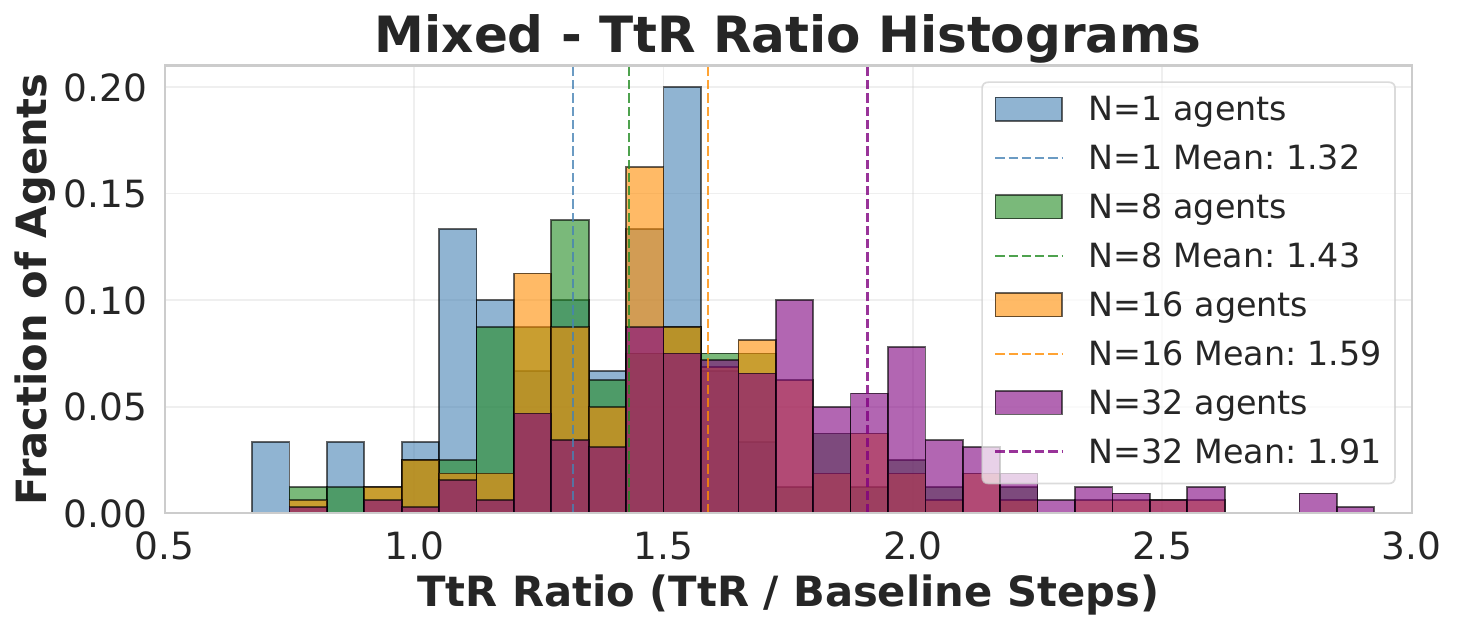}
     \end{subfigure}

\caption{\small
Per-agent delay (from TtR) with \diffplanner~relative to the average time a single agent ($N{=}1$) using \stlpy~takes to satisfy its specification. Histograms summarize 30 runs of the \mixedspec~task with Baseline Steps = 990.
}
\label{fig:delay-per-agent}

	\vspace{-1em}
\end{figure}

Increasing the weight of the achievable loss $\lossacheivable$, which samples the environment to capture multi-agent interactions and \gcbfp~dynamics during planning, lowers TtR by 17.37\% and increases success by 2.37\% in dense settings (32 agents, Table~\ref{tab:mixed-extract-compact-na}).
At this scale, safety-aware STL planning with \diffplanner\ incurs an average per-agent delay (Fig.~\ref{fig:delay-per-agent}) of $1.3$–$1.9\times$ the single-agent \emph{best-case} time reported by \stlpy.

\noindent\textbf{Team Specifications (C1, C2)}
On team tasks (Fig.~\ref{fig:team-spec-results}), \stlpyglobal\ solves allocation near-optimally for \teamredundant~at $N\!=\!8$ but fails to scale (timing out at $N>8$), and per-agent \stlpy\ under oracle allocation (explicit) still suffers safety violations that lower success.
\diffplanner, with no allocation scheme (implicit), holds near $90\%$ average success rate and beats the best \stlpy~baseline by over $30\%$ for $N\!=\!32$.
Thus explicit allocation is sufficient at low $N$, but implicit allocation scales far better as agent safety interactions intensify.

\begin{table}[t]
\centering
\caption{\small
Achievable-loss coefficient ablation.
      D-MA (LA): low coefficient $\coeffacheivable{=}10^{-3}$ (vs.\ $\coeffacheivable{=}1$ for full \diffplanner,  $\coeffstl{=}1$ throughout).
      $\Delta\%$ gives change relative to full \diffplanner. 
}
\begingroup
\setlength{\tabcolsep}{2pt}
\renewcommand{\arraystretch}{0.9}
\scalebox{0.95}{
\resizebox{\columnwidth}{!}{
\begin{tabular}{l|l|cc|cc|cc}
\toprule
\multicolumn{2}{c|}{} & \multicolumn{2}{c|}{Success Rate (\%) $\uparrow$} & \multicolumn{2}{c|}{Planning Time (s) $\downarrow$} & \multicolumn{2}{c}{TtR (steps) $\downarrow$} \\
\multicolumn{2}{c|}{Spec} & {D-MA (LA)} & {$\Delta$\%} & {D-MA (LA)} & {$\Delta$\%} & {D-MA (LA)} & {$\Delta$\%} \\
\midrule
\rotatedHeader[-1em]{3}{\mixedspec} & 8 & 95.00 & -4.60\% & 0.68 & -38.75\% & 1868.17 & +0.98\% \\
 & 16 & 91.25 & -5.81\% & 0.67 & -56.53\% & 1820.72 & +0.46\% \\
 & 32 & 90.00 & -2.37\% & 1.20 & -61.96\% & 2038.92 & +17.37\% \\
\bottomrule
\end{tabular}}
}
\endgroup
\label{tab:mixed-extract-compact-na}
\vspace{-0.1em}
\end{table}

\noindent\textbf{Diversity (C3)} The \divAPC~and \divPO~metrics in Table \ref{tab:results-diversity} show that existing planners (\planner~and \stlpy) tend to generate similar trajectories for all agents, leading to a high overlap in their paths (notably \planner~has over 7 of 16 agents
following close trajectories in the \seq~task and a consistently non-zero \divPO~across tasks).
Our \diffplanner\ planner, on the other hand, generates diverse trajectories for the agents, as indicated by the low %
values (i.e. agents having disjoint paths).

\vspace{-0.5em}
\section{Conclusion}
\label{sec:conclusion}
\vspace{-0.5em}
We propose a novel approach for multi-agent planning that uses STL specification-guided diffusion models to generate joint plans for multiple agents while satisfying safety constraints. 
Our approach is capable of test-time generalization to new specifications with predicates placed anywhere within the trained goal region and is trained on a single agent trajectory dataset, allowing for efficient training and execution on multiple agents. 
Extensive experiments in simulation and real-world environments show that it
matches the scalability of existing learning-based models (viz. \planner) while
outperforming comparable methods in performance and plan diversity.
\vspace{-1em}
\subsection*{Discussion and Outlook}
\label{sec:limitations}

\paragraph{Model-based assumptions}
Our method currently assumes access to accurate system dynamics (as in \citep{zhang_gcbf_2024,qin2021learning}) to deploy the \gcbfp~controller, providing per-step safety via control barrier functions, and to differentiate through $\lossacheivable$ (Sec.~\ref{sec:app-diff}), which quantifies plan achievability with multi-agent interactions. These assumptions 
could be relaxed via model-free techniques \citep{zhang2025dgppo}.%

\paragraph{Interaction with obstacles}
We demonstrate strong scalability primarily in obstacle-free settings (with additional experiments in cluttered environments on our \href{\websiteurl}{website}), showing that the approach has modest hits to success in cluttered environments. Namely, in cluttered maps, coupling \gcbfp~with rich temporal objectives can yield deadlocks under decentralized execution because the barrier function prioritizes instantaneous safety over trajectory-level guarantees \citep{zhang2025dgppo}. This highlights an opportunity to integrate controllers that reason about global or cumulative safety.

\paragraph{Scalability}
Effective coordination beyond 32 agents on fixed maps/predicate sizes likely requires communication or distributed planning. To curb planning-time growth with agent count, faster diffusion inference \citep{li2024faster} or conditional-flow models \citep{meng2025telograf} are promising drop-in replacements, with tunable trade-offs between speed and plan diversity.

\paragraph{Task complexity}
Our focus is on \mastl~and \teamspecs~specifications (per-agent or team-level) with asynchronous timing, aligning with scenarios where users prefer brief delays over collisions (Sec.~\ref{sec:ps-intro}). Extending expressivity to joint-state specifications and synchronized temporal relations (\oldmastl~\citep{Sun2022}), or explicitly modeling delays \citep{yu2023efficient}, presents natural directions for broader applicability.

\balance
\bibliographystyle{IEEEtranN}
\bibliography{refs_ieee_abbrev}
\clearpage
\pagebreak

\end{document}